\documentclass[aps,amsmath,amssymb,twocolumn,10pt,superscriptaddress,prc]{revtex4-1}
\usepackage{graphicx}  
\usepackage[T1]{fontenc}
\usepackage[dvipsnames]{xcolor}
\usepackage[colorlinks=true,linkcolor=blue,citecolor=red,urlcolor=magenta]{hyperref}
\usepackage{charter} 
\usepackage{siunitx}

\newcommand{\CERN}{Physics Department, CERN, 1211 Geneva 23, Switzerland}
\newcommand{\WUT}{\mbox{Faculty of Physics, Warsaw University of Technology, Ulica Koszykowa 75, PL-00662 Warsaw, Poland}}
\newcommand{\IFPAN}{\mbox{Institute of Physics, Polish Academy of Sciences, Aleja Lotnikow 32/46, PL-02668 Warsaw, Poland}}

\begin{document}   

\title{The spin-flip-induced quadrupole resonance in odd-$A$ exotic atoms}

\author{Fredrik P. Gustafsson} 
\affiliation{\CERN}
\author{Daniel P{\k e}cak}
\affiliation{\WUT} \affiliation{\IFPAN} 
\author{Tomasz Sowi\'nski} 
\affiliation{\IFPAN} 

\begin{abstract}
We examine the possible existence of quadrupole resonances in exotic atoms containing odd-$A$ nuclei. We find that the spin-flip of the de-exciting exotic particle can induce a resonance, altering the orbital angular momentum of the nucleus by one quanta. This process results in an excited nucleus with a suppression in x-ray photon emissions during the exotic atom cascade. We provide specific cases of antiprotonic atoms of stable elements where this resonance effect is expected to occur. The study of this phenomena may provide insight into the strong interaction of deeply bound antiprotonic states in the proximity of the unpaired nucleon, and serve as a tool for probing short-lived excited nuclear states.
\end{abstract}  
\maketitle  

\section{Introduction}
Exotic atoms are formed by the substitution of an electron with an exotic negatively charged particle such as an antiproton, muon, pion, or kaon. These exotic particles, owing to their greater mass, are deeply bound within the electron cloud, positioning them in close proximity to the atomic nucleus. This renders them highly susceptible to short-range interactions, such as the weak and strong nuclear forces, making them invaluable tools for probing these fundamental interactions~\cite{2004CohenRPP,1969DevonsAdvNucPhys}. While exotic atoms are typically short-lived, either due to the intrinsic lifetime of the exotic particle or resulting from the annihilation with nucleons in the nucleus spectroscopic analysis of these atoms can offer unique insights into nuclear properties like mass, charge radii, shape and neutron-skin~\cite{1966NuclPhys.87.153, 1966NuclPhys.87.657, 1988NuclPhysA483.619, 1993NuclPhysA561.607,knecht2020study,hori2011two,brown2007neutron}. A notable feature of exotic atoms is their significantly higher transition energies compared to conventional electronic systems. These energies often lie in the range of keV to MeV, as opposed to the eV scale typical of the valence electrons bound in neutral atoms \cite{2008GottaEPJD}. Interestingly, these energies are comparable with energy scales of nuclear phenomena. In the case of an exotic atom containing a negatively charged hadron orbiting the nucleus, such as a negatively charged pion, kaon, or antiproton, the deeply bound states will be directly influenced by the strong nuclear force. Once the orbiting hadron's wavefunction has sufficient overlap with the nucleus, it is rapidly absorbed, hindering the spectroscopic investigation of the deeply bound states that are most sensitive to the strong interaction effects. However, in some cases, the energy of a nuclear quadrupole (E2) excitation can be sufficiently close to that of the transition between states in the exotic atom, resulting in a so-called E2 resonance effect~\cite{1976LeonNuclPhysA}. This resonance effect has been experimentally studied through X-ray spectroscopy of hadronic atoms containing pions, kaons, and antiprotons ~\cite{leon1974e2,batty1978e2,leon1979observation,reidy1985measurements,kanert1986first,1990PhysLettB252.27,1993NuclPhysA561.607, 2004PhysRevC.69.044311}. In these studies, the resonance effect has been used to probe the strong interaction shift and width of the 'hidden' deeply bound state, owing to the quadrupole-induced mixing with a spectroscopically accessible state, revealing information about the strong interaction potential as well as the density distribution of the nuclear periphery. So far, cases of this resonance effect have only been considered for collective quadrupole (E2) excitations of the nucleus, such as rotations and vibrations, leading to a change of the nuclear spin by two quanta, typically favored as the first excited state in even-$A$ nuclei.

In this article, we extend the study of the nuclear resonance effect to odd-$A$ nuclei, with a focus on antiprotonic atoms. We demonstrate that, in this case, the spin-flip of the antiproton may result in the excitation of the nuclear spin by one quanta typically associated with single-particle excitation of the unpaired nucleon. Furthermore, we identify candidates of stable nuclei where we anticipate this spin-flip-induced nuclear resonance effect to occur. Finally, we discuss potential scenarios where this phenomenon could be used as a probe for nuclear structure and strong interaction effects.

It is worth mentioning that spin-flip induced excitations of the nucleus by orbiting particle {\it without} changing their orbital angular momentum was considered previously in the context of heavy odd-$A$ highly charged ions \cite{1998KarpeshinPRC}. A plethora of those effects concerns the mixing of different electronic states and are manifested in spectra \cite{2015KarpeshinNPhysA,2023KarpeshinARXIV,wheeler1949some,karpeshin2006resonance}. Here we extend this approach to cases when heavy orbiting particle, due to the quadrupole interaction with the nucleus, changes also their orbital angular momentum and principal quantum number. In this scenario, the energy released is much larger and can fit to gamma transitions in the nuclear sector.

\section{The system}
The most convenient way of considering corrections coming from the quadrupole interactions is to write the Hamiltonian describing the system of deformed nucleus and orbiting antiproton as
\begin{equation} \label{Ham}
{\cal H} = {\cal H}_{R} + {\cal H}_{r} + {\cal H}_Q,
\end{equation}
where ${\cal H}_{R}$ is the Hamiltonian acting solely in the nucleus subspace while ${\cal H}_{r}$ is the Dirac Hamiltonian for the antiproton in the spherically symmetric electric potential,  $V(r)=-Z/r$. Its bound eigenstates $|njm\kappa\rangle$ are conveniently labeled with the principal quantum number $n$, the total angular momentum $j$, its projection $m$, and the sign of the relativistic quantum number $\kappa=\pm 1$. Although in principle the combined integer $\ell=j + \kappa/2$ is not a good quantum number, for interpretation purposes, it can be viewed as an orbital angular momentum number. In the non-relativistic limit (Pauli approximation), corresponding spinor wavefunctions have a form~\cite{1957BetheSalpeterBook,2006DrakeAMOBook}
\begin{subequations} \label{DiracWave}
\begin{eqnarray}
\langle \boldsymbol{r}|njm+\rangle =& R^n_{j+\text{\textonehalf}}(r)\left(\begin{array}{c}\sqrt{\frac{j-m+1}{2j+2}}\mathrm{Y}_{j+\text{\textonehalf}}^{m-\text{\textonehalf}}(\boldsymbol{\vartheta}) \\ \sqrt{\frac{j+m+1}{2j+2}}\mathrm{Y}_{j+\text{\textonehalf}}^{m+\text{\textonehalf}}(\boldsymbol{\vartheta})\end{array}\right), \\
\langle \boldsymbol{r}|njm-\rangle =& R^n_{j-\text{\textonehalf}}(r)\left(\begin{array}{c}\sqrt{\frac{j+m}{2j}}\mathrm{Y}_{j-\text{\textonehalf}}^{m-\text{\textonehalf}}(\boldsymbol{\vartheta}) \\ -\sqrt{\frac{j-m}{2j}}\mathrm{Y}_{j-\text{\textonehalf}}^{m+\text{\textonehalf}}(\boldsymbol{\vartheta})\end{array}\right),
\end{eqnarray}
\end{subequations}
where $R^n_{\ell}(r)$ encodes radial probability amplitude while $\mathrm{Y}_{\ell}^{m}(\boldsymbol{\vartheta})$ is a spherical harmonic function. In the following, we consider the most typical experimental scenario when the antiproton substitutes one of the innermost electrons, enters the Rydberg-like orbit with the corresponding principal number $n\approx 40$, and then undergoes spontaneous transitions to lower $n$ states emitting x-ray photons~\cite{2004CohenRPP,2008GottaEPJD}. Since Rydberg orbits are spatially well-localized we assume that $\ell$ and $m$ are as large as possible, {\it i.e.}, they are close to the principal quantum number $n$. 

Quantum description of the nucleus is not so straightforward, especially when odd-$A$ nuclei are considered. Therefore we work within the simplest possible framework capturing the essence of the quadrupole resonance phenomena in odd-$A$ systems. This gives us a path to explain in detail all aspects of the phenomena, to specify conditions for its observability, and to estimate its basic measurable features. Of course, one can use more sophisticated nuclear models, especially for cases when quadrupole moments of the nucleus are essentially dependent on its quantum state, to obtain more accurate predictions~\cite{2007PhysRevC.76.054305,otsuka2005evolution,morris2018structure,otsuka2022moments,lechner2023electromagnetic,karthein2023electromagnetic}. Nevertheless, the general scheme outlined below will not change. 

Our framework is based on the fundamental assumption that the density of nuclear matter exhibits minimal dependence on its quantum state and hosts also information about nuclear charge distribution. Assuming additionally that the nucleus is axially symmetric one finds that the deformations of the nucleus are characterized by the intrinsic nuclear quadrupole moment while its rotational quantum states are described within the Bohr-Mottelson collective rotational model framework~\cite{1976BohrRMP,1976MottelsonRMP,1995NaqviZPhysA,2016RowePhysScripta}. Moreover, all the eigenstates of ${\cal H}_R$ are expressed explicitly in terms of Wigner D-matrices 
\begin{multline} \label{CollectiveWave}
\langle\boldsymbol{\Theta}|JKM\rangle =  \sqrt{\frac{2J+1}{16\pi^2}}\Big[\xi_K D^{J}_{MK}(\boldsymbol{\Theta}) \\+ (-1)^{J+K}\xi_{\bar{K}}D^{J}_{M-K}(\boldsymbol{\Theta})\Big].
\end{multline}
Here we assumed that $J$ is half-integer since we focus on odd-$A$ elements, similarly to its projections $K, M$.

\section{Quadrupole interactions}
The remaining part of the Hamiltonian, ${\cal H}_Q$, dominated by the quadrupole interactions, essentially accounts for all contributions not captured by the point-like approximation of the nucleus. Within our framework, this Hamiltonian can be decomposed into terms that act independently on the nucleus and antiproton sectors, as follows~\cite{1976LeonNuclPhysA}
\begin{equation}
{\cal H}_Q = -\frac{2\pi}{5}{\cal Q}_0 \mathrm{Y}_2^0({\boldsymbol \Theta})\, r^{-3}\,\mathrm{Y}_{2}^{0}(\boldsymbol{\vartheta}),
\end{equation}
where ${\cal Q}_0$ is the reduced electric quadrupole transition amplitude which can be extracted from the reduced quadrupole transition probabilities $B(E2)$ measured experimentally \cite{1989RaghavanTables,1993NuclPhysA561.607} or estimated theoretically using different nuclear many-body methods \cite{hergert2020guided,karthein2023electromagnetic}. We measure all energies and lengths in natural units of the Coulomb problem $\alpha^2 m_{\bar{p}}c^2\approx 50\,\mathrm{keV}$ and $\hbar/(\alpha m_{\bar{p}}c)\approx 28.8\,\mathrm{fm}$, while $\alpha\approx 1/137$ is the fine-structure constant. 

It is as long as purely spherically symmetric interaction is considered (vanishing ${\cal H}_Q$), product states of orbiting antiproton $|njm\kappa\rangle$, and nucleus $|JKM\rangle$ diagonalize the Hamiltonian ${\cal H}_0={\cal H}_{R} + {\cal H}_{r}$. Since quadrupole interactions preserve total angular momentum, it is convenient to work in the basis of combined states of well-defined total angular momentum ${\cal F}$ ($|j-J|\leq {\cal F}\leq j+J$) and its projection ${\cal M}$ as
\begin{multline}
|{\cal F}{\cal M};nj\kappa JK\rangle = (-1)^{j+J+{\cal M}}\sqrt{2{\cal F}+1}\\ \times\!\sum_{m=-j}^j \left(\begin{array}{ccc}j & J & {\cal F}\\m & {\cal M}\!-\!m & -{\cal M}\end{array}\right)|njm\kappa\rangle|JK({\cal M}\!-\!m)\rangle.
\end{multline}
In this basis, all the matrix elements of the quadrupole Hamiltonian can be calculated using the decomposition rule~\cite{1966NuclPhys.87.657, 1988NuclPhysA483.619,1988ZareBook}
\begin{multline}\label{TransitionAmplitude}
\langle{\cal F}{\cal M};nj\kappa JK|{\cal H}_Q|{\cal F}'{\cal M}';n'j'\kappa'J'K'\rangle =   \\
-\frac{2\pi}{5}{\cal Q}_0\delta_{\cal M \cal M'}\delta_{\cal F\cal F'} (-1)^{j'+J+\cal F} 
 \left\{\begin{array}{ccc}j & j' & 2\\J' & J & {\cal F}\end{array}\right\} \times
  \\
 \langle JK||\mathrm{Y}_{2}^{0}(\boldsymbol{\Theta})||J'K'\rangle
\langle nj|r^{-3}|n'j'\rangle\langle j\kappa||{\mathrm Y}_{2}^{0}(\boldsymbol{\vartheta})||j'\kappa'\rangle.
\end{multline}
Now it is clear that to obtain the transition amplitudes one can calculate contributing parts independently in nuclear and electronic sectors. However, due to the conservation of total angular momentum $\cal F$ and its projection $\cal M$, final contributions must combine to meet this constraint.

\subsection{Transitions in nuclear sector}

Utilizing explicit expressions for the eigenfunctions of the collective rotational model
\eqref{CollectiveWave} one finds straightforwardly matrix elements of the quadrupole moment operator in the nuclear sector
\begin{multline} \label{MElemNucleus}
\langle JKM|\mathrm{Y}_{2}^{0}(\boldsymbol{\Theta})|J'K'M'\rangle =  \\(-1)^{K-M}\sqrt{\frac{5(2J+1)(2J'+1)}{4\pi}}\\ \times
\left(\begin{array}{ccc}J & 2 & J'\\ -M & 0 & M'\end{array}\right)\left(\begin{array}{ccc}J & 2 & J'\\ -K & 0 & K'\end{array}\right).
\end{multline}
This matrix element is evidently consistent with the Wigner-Eckart theorem and leads us directly to the first desired by the formula \eqref{TransitionAmplitude} expression for the reduced matrix element 
\begin{multline}
\langle JK||\mathrm{Y}_{2}^{0}(\boldsymbol{\Theta})||J'K'\rangle =  \\(-1)^{-J+K}\sqrt{\frac{5(2J+1)(2J'+1)}{4\pi}}
\left(\begin{array}{ccc}J & 2 & J'\\ -K & 0 & K'\end{array}\right).
\end{multline}

Having this expression in hand, one deduces that quadrupole interactions can trigger only a well-specified transition of the nucleus. For example, if the even-$A$ element is considered and the nucleus remains in its ground state $|0^+)$, then the only non-vanishing transition amplitude (equal to $\sqrt{5/(4\pi)} \approx 0.63 $) couples to the first excited state of the same band $|2^+)$. In this particular case the angular momentum of the nucleus increases by two quanta ($\Delta J=2$) and, due to the conservation of ${\cal F}$, it must be compensated by a decrease of the angular momentum of an orbiting particle by the same amount. This is indeed possible since quadrupole electric field inherently couples electronic states of an orbiting particle with $\Delta \ell=2$ (see the next subsection for details).

The situation is slightly different when odd-$A$ elements are considered. Then the nucleus, being a fermionic particle, has non-vanishing half-integer spin with single-particle excitations which typically result in the increase of spin by one quanta. Fortunately, it is quite easy to verify that the matrix element \eqref{MElemNucleus} is also non-zero in these cases. As an example, Ruthenium-101 nucleus with ground state $K=J=5/2^+$ and excited state $J'=7/2^+$ is equal to $\sqrt{25/(7\pi)} \approx 1.07$, {\it i.e.}, is of the same order as the matrix element for even-$A$ nuclei.

\subsection{Transitions of orbiting particle}

Analysis of possible transitions in the orbiting particle sector is similar. By taking the central field solutions of Dirac equation for orbiting antiproton \eqref{DiracWave}, one finds angular matrix elements of the electronic part of the quadrupole interactions
\begin{subequations}
\begin{align} \label{Eq9a}
\langle jm+&|\mathrm{Y}_{20}(\boldsymbol{\vartheta})|j'm'+\rangle = \\
\nonumber
&(-1)^{m-\text{\textonehalf}}\sqrt{\frac{5}{4\pi}}
\delta_{mm'} \left(\begin{array}{ccc}j\!+\!\text{\textonehalf} & 2 & j'\!+\!\text{\textonehalf}\\ 0 & 0 & 0\end{array}\right) \\
\nonumber
&\times\left[\sqrt{(j\!-\!m\!+\!1)(j'\!-\!m\!+\!1)}\left(\begin{array}{ccc}j\!+\!\text{\textonehalf} & 2 & j'\!+\!\text{\textonehalf}\\ -m\!+\!\text{\textonehalf} & 0 & m-\!\text{\textonehalf}\end{array}\right)\right.\\
\nonumber
&- \left.\sqrt{(j\!+\!m\!+\!1)(j'\!+\!m\!+\!1)}\left(\begin{array}{ccc}j\!+\!\text{\textonehalf} & 2 & j'\!+\!\text{\textonehalf}\\ -m\!-\!\text{\textonehalf} & 0 & m\!+\!\text{\textonehalf} \end{array}\right)\right],
\end{align}
%
%
\begin{align} \label{Eq9b}
\langle jm-&|\mathrm{Y}_{20}(\boldsymbol{\vartheta})|j'm'-\rangle = \\
\nonumber
&(-1)^{m-\text{\textonehalf}}\sqrt{\frac{5}{4\pi}}
\delta_{mm'} \left(\begin{array}{ccc}j\!-\!\text{\textonehalf} & 2 & j'\!-\!\text{\textonehalf}\\ 0 & 0 & 0\end{array}\right) \\
\nonumber
&\times\left[\sqrt{(j\!+\!m)(j'\!+\!m)}\left(\begin{array}{ccc}j\!-\!\text{\textonehalf} & 2 & j'\!-\!\text{\textonehalf}\\ -m\!+\!\text{\textonehalf} & 0 & m-\!\text{\textonehalf}\end{array}\right)\right.\\
\nonumber
&- \left.\sqrt{(j\!-\!m)(j'\!-\!m)}\left(\begin{array}{ccc}j\!-\!\text{\textonehalf} & 2 & j'\!-\!\text{\textonehalf}\\ -m\!-\!\text{\textonehalf} & 0 & m\!+\!\text{\textonehalf} \end{array}\right)\right],
\end{align}
%
%
\begin{align} \label{Eq9c}
\langle jm+&|\mathrm{Y}_{20}(\boldsymbol{\vartheta})|j'm'-\rangle = \\
\nonumber
&(-1)^{m+\text{\textonehalf}}\sqrt{\frac{5}{4\pi}}\delta_{mm'}\left(\begin{array}{ccc}j\!+\!\text{\textonehalf} & 2 & j'\!-\!\text{\textonehalf}\\ 0 & 0 & 0\end{array}\right) \\
\nonumber
&\times\left[\sqrt{(j\!-\!m\!+\!1)(j'\!+\!m)}\left(\begin{array}{ccc}j\!+\!\text{\textonehalf} & 2 & j'\!-\!\text{\textonehalf}\\ -m\!+\!\text{\textonehalf} & 0 & m-\!\text{\textonehalf}\end{array}\right)\right.\\
\nonumber
&+ \left.\sqrt{(j\!+\!m\!+\!1)(j'\!-\!m)}\left(\begin{array}{ccc}j\!+\!\text{\textonehalf} & 2 & j'\!-\!\text{\textonehalf}\\ -m\!-\!\text{\textonehalf} & 0 & m\!+\!\text{\textonehalf} \end{array}\right)\right].
\end{align}
\end{subequations}
These matrix elements can be simplified further by exploiting half-integer recursion relations for $3j$-symbols \cite{1968MicuNucPhysA} and their symmetries \cite{1979RaynalJMatPhys,2020PainEPJA}. Then they become manifestly consistent with the Wigner-Eckart theorem
\begin{multline}
\langle jm\kappa|\mathrm{Y}_{20}(\boldsymbol{\vartheta})|j'm'\kappa'\rangle = \\
(-1)^{j-m}\left(\begin{array}{ccc}j & 2 & j'\\ -m & 0 & m'\end{array}\right)\langle j\kappa||\mathrm{Y}_{20}(\boldsymbol{\vartheta})||j'\kappa'\rangle 
\end{multline}
where the reduced matrix elements read
\begin{subequations}
\begin{multline} \label{Eq12a}
\langle j\!+\!||\mathrm{Y}_{20}(\boldsymbol{\vartheta})||j'+\rangle = (-1)^{-j-\text{\textonehalf}}\\
\sqrt{\frac{5(4\!+\!j\!+\!j')(j\!+\!j'\!-\!1)}{4\pi}}\left(\begin{array}{ccc}j\!+\!\text{\textonehalf} & 2 & j'\!+\!\text{\textonehalf}\\ 0 & 0 & 0\end{array}\right)
\end{multline}
\begin{multline}\label{Eq12b}
\langle j\!-\!||\mathrm{Y}_{20}(\boldsymbol{\vartheta})||j'-\rangle = (-1)^{-j+\text{\textonehalf}}\\
\sqrt{\frac{5(3\!+\!j\!+\!j')(j\!+\!j'\!-\!2)}{4\pi}}\left(\begin{array}{ccc}j\!-\!\text{\textonehalf} & 2 & j'\!-\!\text{\textonehalf}\\ 0 & 0 & 0\end{array}\right)
\end{multline}
\begin{multline}\label{Eq12c}
\langle j\!+\!||\mathrm{Y}_{20}(\boldsymbol{\vartheta})||j'-\rangle = (-1)^{-j+\text{\textonehalf}}\\
\sqrt{\frac{5(3\!+\!j\!-\!j')(j'\!-\!j\!+\!2)}{4\pi}}\left(\begin{array}{ccc}j\!+\!\text{\textonehalf} & 2 & j'\!-\!\text{\textonehalf}\\ 0 & 0 & 0\end{array}\right)
\end{multline}
\end{subequations}
\begin{figure}
    \centering
    \includegraphics[width=\linewidth]{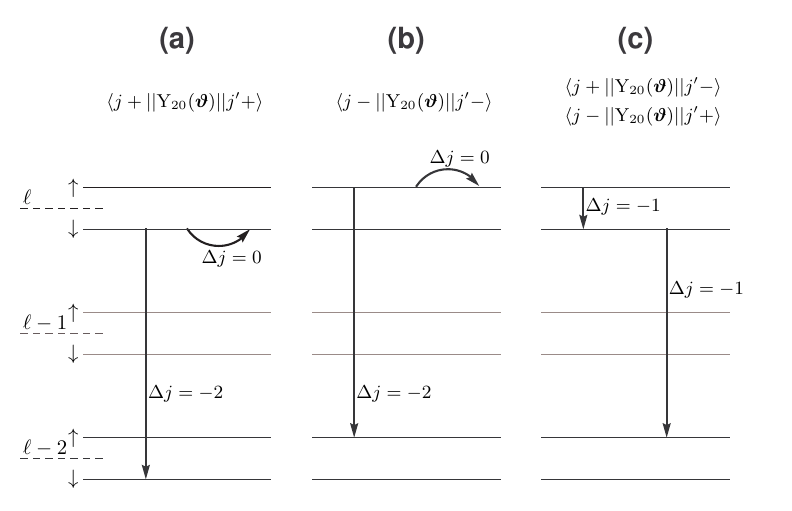}
    \caption{Schematic chart of possible decays from electronic states of an orbiting antiproton triggered by quadrupole interactions, having orbital angular momentum $\ell$. Note that in addition to the typical transitions with $\Delta j=0$ and $\Delta j=-2$ caused by amplitudes \eqref{Eq12a} and \eqref{Eq12b}, transitions that change the total particle spin by $\Delta j=-1$ are also possible due to the simultaneous change of the orbital angular momentum with $\Delta \ell=-2$ and the flip of particle spin. They are controlled by amplitudes \eqref{Eq12c}. }
    \label{Fig1}
\end{figure}

Essentially, there are three different types of non-vanishing matrix elements (schematically presented in Fig.~\ref{Fig1}). The first two involve transitions between states having the same sign of relativistic quantum number $\kappa$. In these cases, the antiproton experiences a small shift of the energy (matrix element with $\Delta j = 0$) or undergoes transition with $\Delta j = -2$ keeping the spin projection unchanged. During this transition, a whole change of the angular momentum comes from the orbital change $\Delta\ell=-2$. As mentioned in the previous subsection, due to the conservation of ${\cal F}$, this transition must be accompanied by a corresponding change of angular momentum of the nucleus, $\Delta J = +2$. For exactly this reason, this transition is typically considered for E2 resonance of even-$A$ nuclei between $|0^+)$ and $|2^+)$ internal states~\cite{1976LeonNuclPhysA,1990PhysLettB252.27,1993NuclPhysA561.607,2004PhysRevC.69.044311}. But of course, it may be also considered in odd-$A$ systems.

The third possibility exists between states of opposite $\kappa$. Then, although the antiproton again undergoes transition with a change of the orbital angular momentum $\Delta\ell=\-2$, the total angular momentum change is $\Delta j=-1$ due to simultaneous flipping of antiproton spin. Note that this transition is also possible without changing orbital angular momentum, $\Delta \ell = 0$, and leads to the mixing of different spin states (see \cite{1998KarpeshinPRC} for details). Correspondingly, due to the conservation of total angular momentum, these transitions must be accompanied by excitations of the nucleus with spin change by one quanta. Thus, the transitions with $\Delta J = 1$ become typical for odd-$A$ elements. It is worth noticing here, that these particular transitions are possible since there is a spin-orbit coupling between the electronic states of the orbiting particle. Thus, its amplitude is evidently smaller when compared to the spin-preserving transitions. However, since in odd-$A$ antiprotonic atoms the spin-preserving channels are locked, this transition is the leading term forced by quadrupole interaction.

From the above discussion it follows that independently of the parity of the nucleus only the transitions with $\Delta \ell=2$ are allowed. Therefore, if we consider a scenario of an exotic particle orbiting in a high Rydberg orbit, {\it i.e.}, $\ell \sim n -1$, the most probable transitions is $|n,\ell=n-1\rangle \rightarrow |n-2,\ell'=n-3\rangle$. This observation is directly related to the fact that the prefactor originating in the radial matrix element $\langle n j|r^{-3}|n' j'\rangle$ quickly decays with increasing difference $\Delta n=n-n'$. Here, one should also note that this factor calculated for the most typical scenario,$\langle nj|r^{-3}|n-2,j-1\rangle$, quickly decays with increasing $n$ (much faster than dipolar factor $\langle nj|r^{-1}|n-1,j-1\rangle$). Therefore, overall quadrupole transition amplitudes with large initial $n$ are strongly suppressed thus we only consider $n\leq 8$ orbitals. 

\section{The resonance}
\begin{figure}
    \centering
    \includegraphics[width=\linewidth]{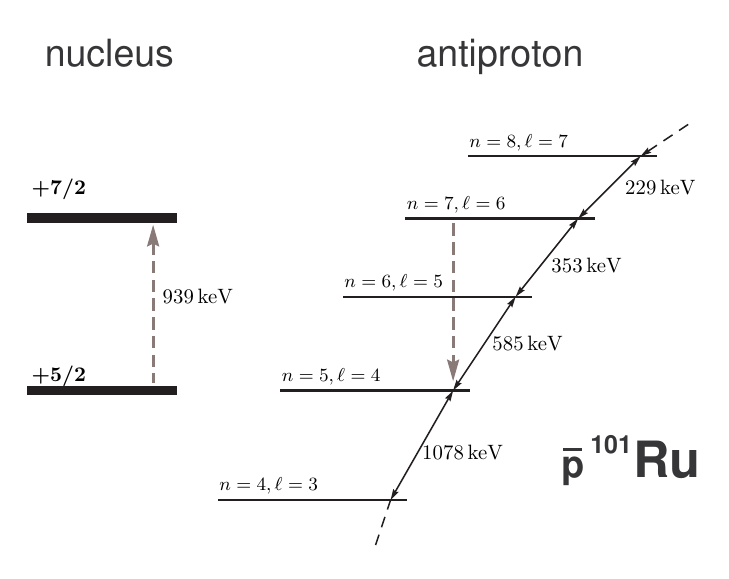}
    \caption{General scheme for matching the energetic condition for quadrupole resonance using the example of an antiprotonic $^{101}$Ru. The $|7/2^+)$ excited state $939\,\mathrm{keV}$ above the $|5/2^+)$ ground state matches the energy difference between electronic states of orbiting antiproton $|n=7,\ell=6\rangle$ and $|n=5,\ell=4\rangle$ (thin dashed arrows). Due to required total angular momentum conservation, the resonance is fully controlled by spin-flipping transitions \eqref{Eq9c}. In this case the states $|7,11/2,+;5/2,K\rangle$ and $|5,9/2,-;7/2,K\rangle$ are resonantly coupled by quadrupole interactions. }
    \label{Fig2}
\end{figure}

The matrix elements of the quadrupole Hamiltonian are rather small when compared with typical energy spacings between coupled electronic states, which is of the order of $\Delta E_{\bar{P}}=E_{nj}-E_{n-2,j-2}$, where $j=\ell\pm 1/2=n-1\pm 1/2$. Thus, the correction is mainly reflected in a relatively small mixing of states in the same (${\cal F},{\cal M}$)-subspace. However, when the energy difference between relevant electronic states of orbiting antiproton is close to the excitation energy of the nucleus, $\Delta E_{N}$, one can suspect resonant behaviour triggered by this coupling. For even-$A$ elements, this behaviour is quite well studied and explored experimentally~\cite{1976LeonNuclPhysA,1988NuclPhysA483.619,1990PhysLettB252.27, 2004PhysRevC.69.044311}. On the contrary, for odd-$A$ elements, due to non-trivial interplay between orbital and spin degrees of motion, observation of the resonance has not yet been explored. An exemplary case of mentioned $^{101}$Ru is displayed in Fig.~\ref{Fig2}. It turns out that the energy difference between $|n=7,\ell=6\rangle$ and $|n=5,\ell=4\rangle$ states is very close to the separation energy between nucleus ground-state $|5/2^+)$ and the $|7/2^+)$ excited state at $939\,\mathrm{keV}$. Moreover, due to the required total angular momentum conservation, this coupling is controlled solely by spin-flipping transitions \eqref{Eq12c} (the two others are forbidden by selection rules) and the initial states $|\mathtt{ini}\rangle=|7,11/2,+;5/2,K\rangle$ are resonantly coupled by quadrupole interactions with final states $|\mathtt{fin}\rangle=|5,9/2,-;7/2,K\rangle$ for any quantum number $K$ that obeys angular momentum projection conservation. When calculated for $K=5/2$ in the maximal ${\cal F}=8$ subspace, the coupling matrix element $\langle \mathtt{fin} |{\cal H}_Q|\mathtt{ini}\rangle/{\cal Q}_0$, is equal to $159\,\mathrm{eV}\mathrm{b}^{-1}$. Thus (provided that unknown ${\cal Q}_0$ is of the order of $1\,\mathrm{b}$) it is of the same order as the energy difference $\Delta E_{N}-\Delta E_{\bar{P}}$ and we expect the resonance effect to be of particular relevance here.

\begin{table*}
\caption{ Examples of stable odd-$A$ antiprotonic atoms where the spin-flip-induced quadrupole resonance is expected. The nuclear transition energies were acquired from \cite{IAEAGammaRayData} and the antiproton transition energies were calculated from the Hamiltonian ${\cal H}_r$. In parenthesis, we indicate tentatively assigned nuclear spins of the excited states. The calculated coupling matrix element $\boldsymbol{\langle \mathtt{fin}|{\cal H}_Q|\mathtt{ini}\rangle/{\cal Q}_0}$ is presented for each scenario. Follow text for more details.}
\centering
\begin{tabular}{ccc|ccS|Sccc}
\hline \hline
{\bf isotope} & 
{\bf \phantom{x}Z\phantom{x}} & 
{\bf \phantom{x}N\phantom{x}} & 
\multicolumn{2}{c}{\bf nuclear spin} & 
$\boldsymbol{\Delta E_{N}}$ & 
$\boldsymbol{\Delta E_{\bar{P}}}$ & 
\multicolumn{2}{c}{$\boldsymbol{n}$} &
$\boldsymbol{\langle \mathtt{fin}|{\cal H}_Q|\mathtt{ini}\rangle/{\cal Q}_0}$
\\ 
&&&
{\bf \phantom{xx}ground\phantom{xx}} & 
{\bf \phantom{xx}excited\phantom{xx}} & 
{\bf (keV)} & 
{\bf (keV)} &
{\bf \phantom{xx}initial\phantom{xx}} & 
{\bf \phantom{xx}final\phantom{xx}}  &
{\bf (eV\,b$^{-1}$)}
\\
\hline \hline
&&& \multicolumn{3}{c|}{\bf NUCLEUS} &
\multicolumn{3}{c}{\bf ANTIPROTON} \\
\hline 
 $^{101}$Ru   & 44 & 57  & 5/2  & (7/2) & 938.65  & 939.40  & 7  & 5 & 159          \\
 $^{111}$Cd & 48 & 63 & 1/2 & 3/2 & 1115.57 & 1119.20 & 7 & 5 &92 \\
 $^{123}$Sb    & 51 & 72  & 7/2  & (9/2)   & 1260.80   & 1264.81   & 7 & 5  &258       \\
 $^{165}$Ho   & 67 & 98 & 7/2  & (9/2)   & 2178.00 & 2189.96   & 7 & 5  &584       \\
 $^{169}$Tm   & 69 & 100 & 1/2  & (3/2)   & 2312.2 & 2323.38   & 7 & 5 &275        \\
 $^{183}$W   & 74 & 109 & 1/2  & (3/2)   & 2667.8 & 2667.47   & 7 & 5 &  339      \\
 $^{203}$Tl   & 81 & 122 & 1/2  & (3/2)   & 1988.88 & 1987.73   & 8 & 6  &159     \\
\hline \hline
\end{tabular}
\label{Table1}
\end{table*}

Let us mention here, that in fact, the scenario discussed above for antiprotonic ruthenium is not unique and in principle may be observed for different stable and unstable isotopes
as long as the conditions of the resonance phenomena are satisfied. In Table.~\ref{Table1} we list candidates of odd-$A$ stable elements for which the energy matching for quadrupole transition is comparable with the corresponding quadrupole coupling matrix element. More precisely, we assume that the effects of quadrupole resonance can be experimentally relevant if the coupling matrix element is not less than $5\%$ of the energy difference between coupled states. Obviously, due to the quick decaying of the quadrupole matrix element $\langle n j|r^{-3}|n-2,j'\rangle$ with increasing $n$, the most interesting are these elements for which the resonant $n$ is small, since then the energy matching may be less accurate. On the other hand, when $n$ is sufficiently small, the corresponding antiprotonic orbit is in close proximity to the surface of the nucleus. In this case, the resonant condition is affected by the influence of the strong interaction which is neglected in these studies.

This spin-flip-induced resonance effect could be revealed by performing x-ray spectroscopy of the antiprotonic atom cascade, by impinging a low energy antiproton beam on a target material. Similar to the observation of the standard E2 resonance effect, we anticipate that in the exemplary case of $^{101}$Ru, the $n=7$ to $n=6$ antiproton x-ray transition will be attenuated due to spin-flip resonance mixing with the deeply bound $n=5$ state, as compared with its neighboring even-$A$ isotopes $^{100}$Ru and $^{102}$Ru. Thus, the spin-flip-induced resonance effect could allow the study of the strong interaction shifts and width of antiproton orbits in the close proximity to the nucleus, which may give further insight into the influence of the unpaired nucleon on the nuclear periphery. The spin-flip resonance effect should thus be taken into account when studying the strong interaction width and shifts from measured x-ray spectra of antiprotonic atom cascades, especially for odd-$A$ nuclei. Furthermore, if the spin and moments of the nuclear ground state are well understood then the observation of this resonance effect may prove itself as a complementary tool for decoupling the properties of the short-lived excited nuclear state. For example, the spin of the upper-level nuclear states of the proposed transitions have only been tentatively assigned, see Table.~\ref{Table1}. Hence, observing this resonance effect could possibly yield insights into the nuclear spin and quadrupole moment of the short-lived excited states, given that an incorrectly assigned nuclear spin of the upper state would preclude the occurrence of this phenomenon. However, in order to decouple the strong interaction effects more detailed calculations are required which is beyond the scope of this work.

\section{Final remarks} 
This work extends the study of the nuclear resonance effect of exotic atoms by focusing on odd-$A$ nuclei with half-integer ground-state spin. We show that $\Delta J=1$ nuclear excitations can be directly triggered by \(\Delta \ell=2\) transitions of the decaying antiproton undergoing spin-flip. We present cases of odd-$A$ antiprotonic atoms where the spin-flip-induced quadrupole resonance effect is expected to occur. The observation of this effect could reveal the strong interaction influence on deeply bound antiproton orbitals on nuclei with unpaired nucleons and potentially serve as a tool for benchmarking the nuclear structure of short-lived excited nuclear states. For completeness, let us also mention that the same spin-flip-induced quadrupole resonance could in principle be observed for highly charged ions or other exotic atoms containing fermionic particles.

The rapid development of new techniques at the ELENA/AD antiproton decelerator facility at CERN, combined with a growing interest in antimatter-matter bound systems, sets the stage for numerous upcoming measurements in the near future \cite{doser2022antiprotonic}. Recent technical developments at the AEgIS experiment within this facility will enable the controlled formation and study of antiprotonic atoms, including the synthesis of highly charged nuclear fragments for precision spectroscopy \cite{rodin2022low,kornakov2023synthesis,AEgISReport2022,AEgISReport2023}. Furthermore, the upcoming availability of antiprotons in portable traps will provide the nuclear physics community with access to antiprotons \cite{aumann2022puma}, facilitating new avenues for nuclear structure studies, such as the one proposed in this work.

\acknowledgments
The authors are very grateful to Michael Doser, Simon Lechner, Piotr Magierski, Leszek Pr\'ochniak, and Katarzyna Wrzosek-Lipska for their insightful comments and inspiring questions. The authors are also indebted for many valuable discussions with AEgIS Collaboration members.
The work by T. S. and D. P. was supported as part of a project funded by the Polish Ministry of Education and Science on the basis of agreement no. 2022/WK/06.

\bibliography{biblio}

\begin{thebibliography}{48}%
\makeatletter
\providecommand \@ifxundefined [1]{%
 \@ifx{#1\undefined}
}%
\providecommand \@ifnum [1]{%
 \ifnum #1\expandafter \@firstoftwo
 \else \expandafter \@secondoftwo
 \fi
}%
\providecommand \@ifx [1]{%
 \ifx #1\expandafter \@firstoftwo
 \else \expandafter \@secondoftwo
 \fi
}%
\providecommand \natexlab [1]{#1}%
\providecommand \enquote  [1]{``#1''}%
\providecommand \bibnamefont  [1]{#1}%
\providecommand \bibfnamefont [1]{#1}%
\providecommand \citenamefont [1]{#1}%
\providecommand \href@noop [0]{\@secondoftwo}%
\providecommand \href [0]{\begingroup \@sanitize@url \@href}%
\providecommand \@href[1]{\@@startlink{#1}\@@href}%
\providecommand \@@href[1]{\endgroup#1\@@endlink}%
\providecommand \@sanitize@url [0]{\catcode `\\12\catcode `\$12\catcode
  `\&12\catcode `\#12\catcode `\^12\catcode `\_12\catcode `\%12\relax}%
\providecommand \@@startlink[1]{}%
\providecommand \@@endlink[0]{}%
\providecommand \url  [0]{\begingroup\@sanitize@url \@url }%
\providecommand \@url [1]{\endgroup\@href {#1}{\urlprefix }}%
\providecommand \urlprefix  [0]{URL }%
\providecommand \Eprint [0]{\href }%
\providecommand \doibase [0]{http://dx.doi.org/}%
\providecommand \selectlanguage [0]{\@gobble}%
\providecommand \bibinfo  [0]{\@secondoftwo}%
\providecommand \bibfield  [0]{\@secondoftwo}%
\providecommand \translation [1]{[#1]}%
\providecommand \BibitemOpen [0]{}%
\providecommand \bibitemStop [0]{}%
\providecommand \bibitemNoStop [0]{.\EOS\space}%
\providecommand \EOS [0]{\spacefactor3000\relax}%
\providecommand \BibitemShut  [1]{\csname bibitem#1\endcsname}%
\let\auto@bib@innerbib\@empty
\bibitem [{\citenamefont {Cohen}(2004)}]{2004CohenRPP}%
  \BibitemOpen
  \bibfield  {author} {\bibinfo {author} {\bibfnamefont {J.~S.}\ \bibnamefont
  {Cohen}},\ }\href {\doibase 10.1088/0034-4885/67/10/R02} {\bibfield
  {journal} {\bibinfo  {journal} {Reports on Progress in Physics}\ }\textbf
  {\bibinfo {volume} {67}},\ \bibinfo {pages} {1769} (\bibinfo {year}
  {2004})}\BibitemShut {NoStop}%
\bibitem [{\citenamefont {Devons}\ and\ \citenamefont
  {Duerdoth}(1969)}]{1969DevonsAdvNucPhys}%
  \BibitemOpen
  \bibfield  {author} {\bibinfo {author} {\bibfnamefont {S.}~\bibnamefont
  {Devons}}\ and\ \bibinfo {author} {\bibfnamefont {I.}~\bibnamefont
  {Duerdoth}},\ }\enquote {\bibinfo {title} {Muonic atoms},}\ in\ \href
  {\doibase 10.1007/978-1-4684-8343-7_5} {\emph {\bibinfo {booktitle} {Advances
  in Nuclear Physics: Volume 2}}},\ \bibinfo {editor} {edited by\ \bibinfo
  {editor} {\bibfnamefont {M.}~\bibnamefont {Baranger}}\ and\ \bibinfo {editor}
  {\bibfnamefont {E.}~\bibnamefont {Vogt}}}\ (\bibinfo  {publisher} {Springer
  US},\ \bibinfo {address} {New York, NY},\ \bibinfo {year} {1969})\ pp.\
  \bibinfo {pages} {295--423}\BibitemShut {NoStop}%
\bibitem [{\citenamefont {Acker}(1966)}]{1966NuclPhys.87.153}%
  \BibitemOpen
  \bibfield  {author} {\bibinfo {author} {\bibfnamefont {H.}~\bibnamefont
  {Acker}},\ }\href {\doibase https://doi.org/10.1016/0029-5582(66)90368-3}
  {\bibfield  {journal} {\bibinfo  {journal} {Nuclear Physics}\ }\textbf
  {\bibinfo {volume} {87}},\ \bibinfo {pages} {153} (\bibinfo {year}
  {1966})}\BibitemShut {NoStop}%
\bibitem [{\citenamefont {{De Wit}}\ \emph {et~al.}(1966)\citenamefont {{De
  Wit}}, \citenamefont {Backenstoss}, \citenamefont {Daum}, \citenamefont
  {Sens},\ and\ \citenamefont {Acker}}]{1966NuclPhys.87.657}%
  \BibitemOpen
  \bibfield  {author} {\bibinfo {author} {\bibfnamefont {S.}~\bibnamefont {{De
  Wit}}}, \bibinfo {author} {\bibfnamefont {G.}~\bibnamefont {Backenstoss}},
  \bibinfo {author} {\bibfnamefont {C.}~\bibnamefont {Daum}}, \bibinfo {author}
  {\bibfnamefont {J.}~\bibnamefont {Sens}}, \ and\ \bibinfo {author}
  {\bibfnamefont {H.}~\bibnamefont {Acker}},\ }\href {\doibase
  https://doi.org/10.1016/0029-5582(67)90003-X} {\bibfield  {journal} {\bibinfo
   {journal} {Nuclear Physics}\ }\textbf {\bibinfo {volume} {87}},\ \bibinfo
  {pages} {657} (\bibinfo {year} {1966})}\BibitemShut {NoStop}%
\bibitem [{\citenamefont {Green}\ \emph {et~al.}(1988)\citenamefont {Green},
  \citenamefont {Liu},\ and\ \citenamefont {Wycech}}]{1988NuclPhysA483.619}%
  \BibitemOpen
  \bibfield  {author} {\bibinfo {author} {\bibfnamefont {A.}~\bibnamefont
  {Green}}, \bibinfo {author} {\bibfnamefont {G.}~\bibnamefont {Liu}}, \ and\
  \bibinfo {author} {\bibfnamefont {S.}~\bibnamefont {Wycech}},\ }\href
  {\doibase https://doi.org/10.1016/0375-9474(88)90087-5} {\bibfield  {journal}
  {\bibinfo  {journal} {Nuclear Physics A}\ }\textbf {\bibinfo {volume}
  {483}},\ \bibinfo {pages} {619} (\bibinfo {year} {1988})}\BibitemShut
  {NoStop}%
\bibitem [{\citenamefont {Wycech}\ \emph {et~al.}(1993)\citenamefont {Wycech},
  \citenamefont {Hartmann}, \citenamefont {Daniel}, \citenamefont {Kanert},
  \citenamefont {Plendl}, \citenamefont {{von Egidy}}, \citenamefont {Reidy},
  \citenamefont {Nicholas}, \citenamefont {Redmond}, \citenamefont {Koch},
  \citenamefont {Kreissl}, \citenamefont {Poth},\ and\ \citenamefont
  {Rohmann}}]{1993NuclPhysA561.607}%
  \BibitemOpen
  \bibfield  {author} {\bibinfo {author} {\bibfnamefont {S.}~\bibnamefont
  {Wycech}}, \bibinfo {author} {\bibfnamefont {F.}~\bibnamefont {Hartmann}},
  \bibinfo {author} {\bibfnamefont {H.}~\bibnamefont {Daniel}}, \bibinfo
  {author} {\bibfnamefont {W.}~\bibnamefont {Kanert}}, \bibinfo {author}
  {\bibfnamefont {H.}~\bibnamefont {Plendl}}, \bibinfo {author} {\bibfnamefont
  {T.}~\bibnamefont {{von Egidy}}}, \bibinfo {author} {\bibfnamefont
  {J.}~\bibnamefont {Reidy}}, \bibinfo {author} {\bibfnamefont
  {M.}~\bibnamefont {Nicholas}}, \bibinfo {author} {\bibfnamefont
  {L.}~\bibnamefont {Redmond}}, \bibinfo {author} {\bibfnamefont
  {H.}~\bibnamefont {Koch}}, \bibinfo {author} {\bibfnamefont {A.}~\bibnamefont
  {Kreissl}}, \bibinfo {author} {\bibfnamefont {H.}~\bibnamefont {Poth}}, \
  and\ \bibinfo {author} {\bibfnamefont {D.}~\bibnamefont {Rohmann}},\ }\href
  {\doibase https://doi.org/10.1016/0375-9474(93)90068-9} {\bibfield  {journal}
  {\bibinfo  {journal} {Nuclear Physics A}\ }\textbf {\bibinfo {volume}
  {561}},\ \bibinfo {pages} {607} (\bibinfo {year} {1993})}\BibitemShut
  {NoStop}%
\bibitem [{\citenamefont {Knecht}\ \emph {et~al.}(2020)\citenamefont {Knecht},
  \citenamefont {Skawran},\ and\ \citenamefont {Vogiatzi}}]{knecht2020study}%
  \BibitemOpen
  \bibfield  {author} {\bibinfo {author} {\bibfnamefont {A.}~\bibnamefont
  {Knecht}}, \bibinfo {author} {\bibfnamefont {A.}~\bibnamefont {Skawran}}, \
  and\ \bibinfo {author} {\bibfnamefont {S.~M.}\ \bibnamefont {Vogiatzi}},\
  }\href@noop {} {\bibfield  {journal} {\bibinfo  {journal} {The European
  Physical Journal Plus}\ }\textbf {\bibinfo {volume} {135}},\ \bibinfo {pages}
  {777} (\bibinfo {year} {2020})}\BibitemShut {NoStop}%
\bibitem [{\citenamefont {Hori}\ \emph {et~al.}(2011)\citenamefont {Hori},
  \citenamefont {S{\'o}t{\'e}r}, \citenamefont {Barna}, \citenamefont {Dax},
  \citenamefont {Hayano}, \citenamefont {Friedreich}, \citenamefont
  {Juh{\'a}sz}, \citenamefont {Pask}, \citenamefont {Widmann}, \citenamefont
  {Horv{\'a}th} \emph {et~al.}}]{hori2011two}%
  \BibitemOpen
  \bibfield  {author} {\bibinfo {author} {\bibfnamefont {M.}~\bibnamefont
  {Hori}}, \bibinfo {author} {\bibfnamefont {A.}~\bibnamefont {S{\'o}t{\'e}r}},
  \bibinfo {author} {\bibfnamefont {D.}~\bibnamefont {Barna}}, \bibinfo
  {author} {\bibfnamefont {A.}~\bibnamefont {Dax}}, \bibinfo {author}
  {\bibfnamefont {R.}~\bibnamefont {Hayano}}, \bibinfo {author} {\bibfnamefont
  {S.}~\bibnamefont {Friedreich}}, \bibinfo {author} {\bibfnamefont
  {B.}~\bibnamefont {Juh{\'a}sz}}, \bibinfo {author} {\bibfnamefont
  {T.}~\bibnamefont {Pask}}, \bibinfo {author} {\bibfnamefont {E.}~\bibnamefont
  {Widmann}}, \bibinfo {author} {\bibfnamefont {D.}~\bibnamefont
  {Horv{\'a}th}},  \emph {et~al.},\ }\href@noop {} {\bibfield  {journal}
  {\bibinfo  {journal} {Nature}\ }\textbf {\bibinfo {volume} {475}},\ \bibinfo
  {pages} {484} (\bibinfo {year} {2011})}\BibitemShut {NoStop}%
\bibitem [{\citenamefont {Brown}\ \emph {et~al.}(2007)\citenamefont {Brown},
  \citenamefont {Shen}, \citenamefont {Hillhouse}, \citenamefont {Meng},\ and\
  \citenamefont {Trzci{\'n}ska}}]{brown2007neutron}%
  \BibitemOpen
  \bibfield  {author} {\bibinfo {author} {\bibfnamefont {B.~A.}\ \bibnamefont
  {Brown}}, \bibinfo {author} {\bibfnamefont {G.}~\bibnamefont {Shen}},
  \bibinfo {author} {\bibfnamefont {G.}~\bibnamefont {Hillhouse}}, \bibinfo
  {author} {\bibfnamefont {J.}~\bibnamefont {Meng}}, \ and\ \bibinfo {author}
  {\bibfnamefont {A.}~\bibnamefont {Trzci{\'n}ska}},\ }\href@noop {} {\bibfield
   {journal} {\bibinfo  {journal} {Physical Review C}\ }\textbf {\bibinfo
  {volume} {76}},\ \bibinfo {pages} {034305} (\bibinfo {year}
  {2007})}\BibitemShut {NoStop}%
\bibitem [{\citenamefont {Gotta}\ \emph {et~al.}(2008)\citenamefont {Gotta},
  \citenamefont {Rashid}, \citenamefont {Fricke}, \citenamefont {Indelicato},\
  and\ \citenamefont {Simons}}]{2008GottaEPJD}%
  \BibitemOpen
  \bibfield  {author} {\bibinfo {author} {\bibfnamefont {D.}~\bibnamefont
  {Gotta}}, \bibinfo {author} {\bibfnamefont {K.}~\bibnamefont {Rashid}},
  \bibinfo {author} {\bibfnamefont {B.}~\bibnamefont {Fricke}}, \bibinfo
  {author} {\bibfnamefont {P.}~\bibnamefont {Indelicato}}, \ and\ \bibinfo
  {author} {\bibfnamefont {L.~M.}\ \bibnamefont {Simons}},\ }\href {\doibase
  10.1140/epjd/e2008-00025-3} {\bibfield  {journal} {\bibinfo  {journal} {The
  European Physical Journal D}\ }\textbf {\bibinfo {volume} {47}},\ \bibinfo
  {pages} {11} (\bibinfo {year} {2008})}\BibitemShut {NoStop}%
\bibitem [{\citenamefont {Leon}(1976)}]{1976LeonNuclPhysA}%
  \BibitemOpen
  \bibfield  {author} {\bibinfo {author} {\bibfnamefont {M.}~\bibnamefont
  {Leon}},\ }\href {\doibase https://doi.org/10.1016/0375-9474(76)90057-9}
  {\bibfield  {journal} {\bibinfo  {journal} {Nuclear Physics A}\ }\textbf
  {\bibinfo {volume} {260}},\ \bibinfo {pages} {461} (\bibinfo {year}
  {1976})}\BibitemShut {NoStop}%
\bibitem [{\citenamefont {Leon}(1974)}]{leon1974e2}%
  \BibitemOpen
  \bibfield  {author} {\bibinfo {author} {\bibfnamefont {M.}~\bibnamefont
  {Leon}},\ }\href@noop {} {\bibfield  {journal} {\bibinfo  {journal} {Physics
  Letters B}\ }\textbf {\bibinfo {volume} {53}},\ \bibinfo {pages} {141}
  (\bibinfo {year} {1974})}\BibitemShut {NoStop}%
\bibitem [{\citenamefont {Batty}\ \emph {et~al.}(1978)\citenamefont {Batty},
  \citenamefont {Biagi}, \citenamefont {Blecher}, \citenamefont {Kunselman},
  \citenamefont {Riddle}, \citenamefont {Roberts}, \citenamefont {Davies},
  \citenamefont {Pyle}, \citenamefont {Squier}, \citenamefont {Asbury} \emph
  {et~al.}}]{batty1978e2}%
  \BibitemOpen
  \bibfield  {author} {\bibinfo {author} {\bibfnamefont {C.}~\bibnamefont
  {Batty}}, \bibinfo {author} {\bibfnamefont {S.}~\bibnamefont {Biagi}},
  \bibinfo {author} {\bibfnamefont {M.}~\bibnamefont {Blecher}}, \bibinfo
  {author} {\bibfnamefont {R.}~\bibnamefont {Kunselman}}, \bibinfo {author}
  {\bibfnamefont {R.}~\bibnamefont {Riddle}}, \bibinfo {author} {\bibfnamefont
  {B.}~\bibnamefont {Roberts}}, \bibinfo {author} {\bibfnamefont
  {J.}~\bibnamefont {Davies}}, \bibinfo {author} {\bibfnamefont
  {G.}~\bibnamefont {Pyle}}, \bibinfo {author} {\bibfnamefont {G.}~\bibnamefont
  {Squier}}, \bibinfo {author} {\bibfnamefont {D.}~\bibnamefont {Asbury}},
  \emph {et~al.},\ }\href@noop {} {\bibfield  {journal} {\bibinfo  {journal}
  {Nuclear Physics A}\ }\textbf {\bibinfo {volume} {296}},\ \bibinfo {pages}
  {361} (\bibinfo {year} {1978})}\BibitemShut {NoStop}%
\bibitem [{\citenamefont {Leon}\ \emph {et~al.}(1979)\citenamefont {Leon},
  \citenamefont {Bradbury}, \citenamefont {Gram}, \citenamefont {Hutson},
  \citenamefont {Schillaci}, \citenamefont {Hargrove},\ and\ \citenamefont
  {Reidy}}]{leon1979observation}%
  \BibitemOpen
  \bibfield  {author} {\bibinfo {author} {\bibfnamefont {M.}~\bibnamefont
  {Leon}}, \bibinfo {author} {\bibfnamefont {J.}~\bibnamefont {Bradbury}},
  \bibinfo {author} {\bibfnamefont {P.}~\bibnamefont {Gram}}, \bibinfo {author}
  {\bibfnamefont {R.}~\bibnamefont {Hutson}}, \bibinfo {author} {\bibfnamefont
  {M.}~\bibnamefont {Schillaci}}, \bibinfo {author} {\bibfnamefont
  {C.}~\bibnamefont {Hargrove}}, \ and\ \bibinfo {author} {\bibfnamefont
  {J.}~\bibnamefont {Reidy}},\ }\href@noop {} {\bibfield  {journal} {\bibinfo
  {journal} {Nuclear Physics A}\ }\textbf {\bibinfo {volume} {322}},\ \bibinfo
  {pages} {397} (\bibinfo {year} {1979})}\BibitemShut {NoStop}%
\bibitem [{\citenamefont {Reidy}\ \emph {et~al.}(1985)\citenamefont {Reidy},
  \citenamefont {Nicholas}, \citenamefont {Bradbury}, \citenamefont {Gram},
  \citenamefont {Hutson}, \citenamefont {Leon}, \citenamefont {Schillaci},
  \citenamefont {Hartmann},\ and\ \citenamefont
  {Kunselman}}]{reidy1985measurements}%
  \BibitemOpen
  \bibfield  {author} {\bibinfo {author} {\bibfnamefont {J.}~\bibnamefont
  {Reidy}}, \bibinfo {author} {\bibfnamefont {M.}~\bibnamefont {Nicholas}},
  \bibinfo {author} {\bibfnamefont {J.}~\bibnamefont {Bradbury}}, \bibinfo
  {author} {\bibfnamefont {P.}~\bibnamefont {Gram}}, \bibinfo {author}
  {\bibfnamefont {R.}~\bibnamefont {Hutson}}, \bibinfo {author} {\bibfnamefont
  {M.}~\bibnamefont {Leon}}, \bibinfo {author} {\bibfnamefont {M.}~\bibnamefont
  {Schillaci}}, \bibinfo {author} {\bibfnamefont {F.}~\bibnamefont {Hartmann}},
  \ and\ \bibinfo {author} {\bibfnamefont {A.}~\bibnamefont {Kunselman}},\
  }\href@noop {} {\bibfield  {journal} {\bibinfo  {journal} {Physical Review
  C}\ }\textbf {\bibinfo {volume} {32}},\ \bibinfo {pages} {1646} (\bibinfo
  {year} {1985})}\BibitemShut {NoStop}%
\bibitem [{\citenamefont {Kanert}\ \emph {et~al.}(1986)\citenamefont {Kanert},
  \citenamefont {Hartmann}, \citenamefont {Daniel}, \citenamefont {Moser},
  \citenamefont {Schmidt}, \citenamefont {von Egidy}, \citenamefont {Reidy},
  \citenamefont {Nicholas}, \citenamefont {Leon}, \citenamefont {Poth} \emph
  {et~al.}}]{kanert1986first}%
  \BibitemOpen
  \bibfield  {author} {\bibinfo {author} {\bibfnamefont {W.}~\bibnamefont
  {Kanert}}, \bibinfo {author} {\bibfnamefont {F.}~\bibnamefont {Hartmann}},
  \bibinfo {author} {\bibfnamefont {H.}~\bibnamefont {Daniel}}, \bibinfo
  {author} {\bibfnamefont {E.}~\bibnamefont {Moser}}, \bibinfo {author}
  {\bibfnamefont {G.}~\bibnamefont {Schmidt}}, \bibinfo {author} {\bibfnamefont
  {T.}~\bibnamefont {von Egidy}}, \bibinfo {author} {\bibfnamefont
  {J.}~\bibnamefont {Reidy}}, \bibinfo {author} {\bibfnamefont
  {M.}~\bibnamefont {Nicholas}}, \bibinfo {author} {\bibfnamefont
  {M.}~\bibnamefont {Leon}}, \bibinfo {author} {\bibfnamefont {H.}~\bibnamefont
  {Poth}},  \emph {et~al.},\ }\href@noop {} {\bibfield  {journal} {\bibinfo
  {journal} {Physical Review Letters}\ }\textbf {\bibinfo {volume} {56}},\
  \bibinfo {pages} {2368} (\bibinfo {year} {1986})}\BibitemShut {NoStop}%
\bibitem [{\citenamefont {Wycech}\ \emph {et~al.}(1990)\citenamefont {Wycech},
  \citenamefont {Liu},\ and\ \citenamefont {Green}}]{1990PhysLettB252.27}%
  \BibitemOpen
  \bibfield  {author} {\bibinfo {author} {\bibfnamefont {S.}~\bibnamefont
  {Wycech}}, \bibinfo {author} {\bibfnamefont {G.}~\bibnamefont {Liu}}, \ and\
  \bibinfo {author} {\bibfnamefont {A.}~\bibnamefont {Green}},\ }\href
  {\doibase https://doi.org/10.1016/0370-2693(90)91074-L} {\bibfield  {journal}
  {\bibinfo  {journal} {Physics Letters B}\ }\textbf {\bibinfo {volume}
  {252}},\ \bibinfo {pages} {27} (\bibinfo {year} {1990})}\BibitemShut
  {NoStop}%
\bibitem [{\citenamefont {K\l{}os}\ \emph {et~al.}(2004)\citenamefont
  {K\l{}os}, \citenamefont {Wycech}, \citenamefont
  {Trzci\ifmmode~\acute{n}\else \'{n}\fi{}ska}, \citenamefont
  {Jastrz\ifmmode~\mbox{\c{e}}\else \c{e}\fi{}bski}, \citenamefont {Czosnyka},
  \citenamefont {Kisieli\ifmmode~\acute{n}\else \'{n}\fi{}ski}, \citenamefont
  {Lubi\ifmmode~\acute{n}\else \'{n}\fi{}ski}, \citenamefont {Napiorkowski},
  \citenamefont {Pie\ifmmode~\acute{n}\else \'{n}\fi{}kowski}, \citenamefont
  {Hartmann}, \citenamefont {Ketzer}, \citenamefont {Schmidt}, \citenamefont
  {von Egidy}, \citenamefont {Cugnon}, \citenamefont {Gulda}, \citenamefont
  {Kurcewicz},\ and\ \citenamefont {Widmann}}]{2004PhysRevC.69.044311}%
  \BibitemOpen
  \bibfield  {author} {\bibinfo {author} {\bibfnamefont {B.}~\bibnamefont
  {K\l{}os}}, \bibinfo {author} {\bibfnamefont {S.}~\bibnamefont {Wycech}},
  \bibinfo {author} {\bibfnamefont {A.}~\bibnamefont
  {Trzci\ifmmode~\acute{n}\else \'{n}\fi{}ska}}, \bibinfo {author}
  {\bibfnamefont {J.}~\bibnamefont {Jastrz\ifmmode~\mbox{\c{e}}\else
  \c{e}\fi{}bski}}, \bibinfo {author} {\bibfnamefont {T.}~\bibnamefont
  {Czosnyka}}, \bibinfo {author} {\bibfnamefont {M.}~\bibnamefont
  {Kisieli\ifmmode~\acute{n}\else \'{n}\fi{}ski}}, \bibinfo {author}
  {\bibfnamefont {P.}~\bibnamefont {Lubi\ifmmode~\acute{n}\else
  \'{n}\fi{}ski}}, \bibinfo {author} {\bibfnamefont {P.}~\bibnamefont
  {Napiorkowski}}, \bibinfo {author} {\bibfnamefont {L.}~\bibnamefont
  {Pie\ifmmode~\acute{n}\else \'{n}\fi{}kowski}}, \bibinfo {author}
  {\bibfnamefont {F.~J.}\ \bibnamefont {Hartmann}}, \bibinfo {author}
  {\bibfnamefont {B.}~\bibnamefont {Ketzer}}, \bibinfo {author} {\bibfnamefont
  {R.}~\bibnamefont {Schmidt}}, \bibinfo {author} {\bibfnamefont
  {T.}~\bibnamefont {von Egidy}}, \bibinfo {author} {\bibfnamefont
  {J.}~\bibnamefont {Cugnon}}, \bibinfo {author} {\bibfnamefont
  {K.}~\bibnamefont {Gulda}}, \bibinfo {author} {\bibfnamefont
  {W.}~\bibnamefont {Kurcewicz}}, \ and\ \bibinfo {author} {\bibfnamefont
  {E.}~\bibnamefont {Widmann}},\ }\href {\doibase 10.1103/PhysRevC.69.044311}
  {\bibfield  {journal} {\bibinfo  {journal} {Phys. Rev. C}\ }\textbf {\bibinfo
  {volume} {69}},\ \bibinfo {pages} {044311} (\bibinfo {year}
  {2004})}\BibitemShut {NoStop}%
\bibitem [{\citenamefont {Karpeshin}\ \emph {et~al.}(1998)\citenamefont
  {Karpeshin}, \citenamefont {Wycech}, \citenamefont {Band}, \citenamefont
  {Trzhaskovskaya}, \citenamefont {Pf\"utzner},\ and\ \citenamefont
  {\ifmmode~\dot{Z}\else \.{Z}\fi{}ylicz}}]{1998KarpeshinPRC}%
  \BibitemOpen
  \bibfield  {author} {\bibinfo {author} {\bibfnamefont {F.~F.}\ \bibnamefont
  {Karpeshin}}, \bibinfo {author} {\bibfnamefont {S.}~\bibnamefont {Wycech}},
  \bibinfo {author} {\bibfnamefont {I.~M.}\ \bibnamefont {Band}}, \bibinfo
  {author} {\bibfnamefont {M.~B.}\ \bibnamefont {Trzhaskovskaya}}, \bibinfo
  {author} {\bibfnamefont {M.}~\bibnamefont {Pf\"utzner}}, \ and\ \bibinfo
  {author} {\bibfnamefont {J.}~\bibnamefont {\ifmmode~\dot{Z}\else
  \.{Z}\fi{}ylicz}},\ }\href {\doibase 10.1103/PhysRevC.57.3085} {\bibfield
  {journal} {\bibinfo  {journal} {Phys. Rev. C}\ }\textbf {\bibinfo {volume}
  {57}},\ \bibinfo {pages} {3085} (\bibinfo {year} {1998})}\BibitemShut
  {NoStop}%
\bibitem [{\citenamefont {Karpeshin}\ and\ \citenamefont
  {Trzhaskovskaya}(2015)}]{2015KarpeshinNPhysA}%
  \BibitemOpen
  \bibfield  {author} {\bibinfo {author} {\bibfnamefont {F.}~\bibnamefont
  {Karpeshin}}\ and\ \bibinfo {author} {\bibfnamefont {M.}~\bibnamefont
  {Trzhaskovskaya}},\ }\href {\doibase
  https://doi.org/10.1016/j.nuclphysa.2015.06.001} {\bibfield  {journal}
  {\bibinfo  {journal} {Nuclear Physics A}\ }\textbf {\bibinfo {volume}
  {941}},\ \bibinfo {pages} {66} (\bibinfo {year} {2015})}\BibitemShut
  {NoStop}%
\bibitem [{\citenamefont {Karpeshin}(2023)}]{2023KarpeshinARXIV}%
  \BibitemOpen
  \bibfield  {author} {\bibinfo {author} {\bibfnamefont {F.~F.}\ \bibnamefont
  {Karpeshin}},\ }\href@noop {} {\enquote {\bibinfo {title} {Revision of
  analytical properties of reaction amplitude near thresholds using the example
  of muon-induced prompt fission},}\ } (\bibinfo {year} {2023}),\ \Eprint
  {http://arxiv.org/abs/2310.19421} {arXiv:2310.19421} \BibitemShut {NoStop}%
\bibitem [{\citenamefont {Wheeler}(1949)}]{wheeler1949some}%
  \BibitemOpen
  \bibfield  {author} {\bibinfo {author} {\bibfnamefont {J.~A.}\ \bibnamefont
  {Wheeler}},\ }\href@noop {} {\bibfield  {journal} {\bibinfo  {journal}
  {Reviews of Modern Physics}\ }\textbf {\bibinfo {volume} {21}},\ \bibinfo
  {pages} {133} (\bibinfo {year} {1949})}\BibitemShut {NoStop}%
\bibitem [{\citenamefont {Karpeshin}(2006)}]{karpeshin2006resonance}%
  \BibitemOpen
  \bibfield  {author} {\bibinfo {author} {\bibfnamefont {F.}~\bibnamefont
  {Karpeshin}},\ }\href@noop {} {\bibfield  {journal} {\bibinfo  {journal}
  {Physics of Particles and Nuclei}\ }\textbf {\bibinfo {volume} {37}},\
  \bibinfo {pages} {284} (\bibinfo {year} {2006})}\BibitemShut {NoStop}%
\bibitem [{\citenamefont {Bethe}\ and\ \citenamefont
  {Salpeter}(1957)}]{1957BetheSalpeterBook}%
  \BibitemOpen
  \bibfield  {author} {\bibinfo {author} {\bibfnamefont {H.}~\bibnamefont
  {Bethe}}\ and\ \bibinfo {author} {\bibfnamefont {E.}~\bibnamefont
  {Salpeter}},\ }\href {\doibase https://doi.org/10.1007/978-3-662-12869-5}
  {\emph {\bibinfo {title} {Quantum Mechanics of One- and Two-Electron
  Atoms}}}\ (\bibinfo  {publisher} {Springer, Berlin},\ \bibinfo {year}
  {1957})\BibitemShut {NoStop}%
\bibitem [{\citenamefont {Drake}(2006)}]{2006DrakeAMOBook}%
  \BibitemOpen
  \bibfield  {author} {\bibinfo {author} {\bibfnamefont {G.}~\bibnamefont
  {Drake}},\ }\href {\doibase https://doi.org/10.1007/978-0-387-26308-3} {\emph
  {\bibinfo {title} {Handbook of Atomic, Molecular, and Optical Physics}}}\
  (\bibinfo  {publisher} {Springer, New York},\ \bibinfo {year}
  {2006})\BibitemShut {NoStop}%
\bibitem [{\citenamefont {Jia}\ \emph {et~al.}(2007)\citenamefont {Jia},
  \citenamefont {Zhang},\ and\ \citenamefont {Zhao}}]{2007PhysRevC.76.054305}%
  \BibitemOpen
  \bibfield  {author} {\bibinfo {author} {\bibfnamefont {L.~Y.}\ \bibnamefont
  {Jia}}, \bibinfo {author} {\bibfnamefont {H.}~\bibnamefont {Zhang}}, \ and\
  \bibinfo {author} {\bibfnamefont {Y.~M.}\ \bibnamefont {Zhao}},\ }\href
  {\doibase 10.1103/PhysRevC.76.054305} {\bibfield  {journal} {\bibinfo
  {journal} {Phys. Rev. C}\ }\textbf {\bibinfo {volume} {76}},\ \bibinfo
  {pages} {054305} (\bibinfo {year} {2007})}\BibitemShut {NoStop}%
\bibitem [{\citenamefont {Otsuka}\ \emph {et~al.}(2005)\citenamefont {Otsuka},
  \citenamefont {Suzuki}, \citenamefont {Fujimoto}, \citenamefont {Grawe},\
  and\ \citenamefont {Akaishi}}]{otsuka2005evolution}%
  \BibitemOpen
  \bibfield  {author} {\bibinfo {author} {\bibfnamefont {T.}~\bibnamefont
  {Otsuka}}, \bibinfo {author} {\bibfnamefont {T.}~\bibnamefont {Suzuki}},
  \bibinfo {author} {\bibfnamefont {R.}~\bibnamefont {Fujimoto}}, \bibinfo
  {author} {\bibfnamefont {H.}~\bibnamefont {Grawe}}, \ and\ \bibinfo {author}
  {\bibfnamefont {Y.}~\bibnamefont {Akaishi}},\ }\href@noop {} {\bibfield
  {journal} {\bibinfo  {journal} {Physical Review Letters}\ }\textbf {\bibinfo
  {volume} {95}},\ \bibinfo {pages} {232502} (\bibinfo {year}
  {2005})}\BibitemShut {NoStop}%
\bibitem [{\citenamefont {Morris}\ \emph {et~al.}(2018)\citenamefont {Morris},
  \citenamefont {Simonis}, \citenamefont {Stroberg}, \citenamefont {Stumpf},
  \citenamefont {Hagen}, \citenamefont {Holt}, \citenamefont {Jansen},
  \citenamefont {Papenbrock}, \citenamefont {Roth},\ and\ \citenamefont
  {Schwenk}}]{morris2018structure}%
  \BibitemOpen
  \bibfield  {author} {\bibinfo {author} {\bibfnamefont {T.~D.}\ \bibnamefont
  {Morris}}, \bibinfo {author} {\bibfnamefont {J.}~\bibnamefont {Simonis}},
  \bibinfo {author} {\bibfnamefont {S.}~\bibnamefont {Stroberg}}, \bibinfo
  {author} {\bibfnamefont {C.}~\bibnamefont {Stumpf}}, \bibinfo {author}
  {\bibfnamefont {G.}~\bibnamefont {Hagen}}, \bibinfo {author} {\bibfnamefont
  {J.}~\bibnamefont {Holt}}, \bibinfo {author} {\bibfnamefont {G.~R.}\
  \bibnamefont {Jansen}}, \bibinfo {author} {\bibfnamefont {T.}~\bibnamefont
  {Papenbrock}}, \bibinfo {author} {\bibfnamefont {R.}~\bibnamefont {Roth}}, \
  and\ \bibinfo {author} {\bibfnamefont {A.}~\bibnamefont {Schwenk}},\
  }\href@noop {} {\bibfield  {journal} {\bibinfo  {journal} {Physical review
  letters}\ }\textbf {\bibinfo {volume} {120}},\ \bibinfo {pages} {152503}
  (\bibinfo {year} {2018})}\BibitemShut {NoStop}%
\bibitem [{\citenamefont {Otsuka}\ \emph {et~al.}(2022)\citenamefont {Otsuka},
  \citenamefont {Shimizu},\ and\ \citenamefont {Tsunoda}}]{otsuka2022moments}%
  \BibitemOpen
  \bibfield  {author} {\bibinfo {author} {\bibfnamefont {T.}~\bibnamefont
  {Otsuka}}, \bibinfo {author} {\bibfnamefont {N.}~\bibnamefont {Shimizu}}, \
  and\ \bibinfo {author} {\bibfnamefont {Y.}~\bibnamefont {Tsunoda}},\
  }\href@noop {} {\bibfield  {journal} {\bibinfo  {journal} {Physical Review
  C}\ }\textbf {\bibinfo {volume} {105}},\ \bibinfo {pages} {014319} (\bibinfo
  {year} {2022})}\BibitemShut {NoStop}%
\bibitem [{\citenamefont {Lechner}\ \emph {et~al.}(2023)\citenamefont
  {Lechner}, \citenamefont {Miyagi}, \citenamefont {Xu}, \citenamefont
  {Bissell}, \citenamefont {Blaum}, \citenamefont {Cheal}, \citenamefont
  {Devlin}, \citenamefont {Garcia~Ruiz}, \citenamefont {Ginges}, \citenamefont
  {Heylen} \emph {et~al.}}]{lechner2023electromagnetic}%
  \BibitemOpen
  \bibfield  {author} {\bibinfo {author} {\bibfnamefont {S.}~\bibnamefont
  {Lechner}}, \bibinfo {author} {\bibfnamefont {T.}~\bibnamefont {Miyagi}},
  \bibinfo {author} {\bibfnamefont {Z.}~\bibnamefont {Xu}}, \bibinfo {author}
  {\bibfnamefont {M.}~\bibnamefont {Bissell}}, \bibinfo {author} {\bibfnamefont
  {K.}~\bibnamefont {Blaum}}, \bibinfo {author} {\bibfnamefont
  {B.}~\bibnamefont {Cheal}}, \bibinfo {author} {\bibfnamefont
  {C.}~\bibnamefont {Devlin}}, \bibinfo {author} {\bibfnamefont
  {R.}~\bibnamefont {Garcia~Ruiz}}, \bibinfo {author} {\bibfnamefont
  {J.}~\bibnamefont {Ginges}}, \bibinfo {author} {\bibfnamefont
  {H.}~\bibnamefont {Heylen}},  \emph {et~al.},\ }\href@noop {} {\bibfield
  {journal} {\bibinfo  {journal} {Physics Letters B}\ }\textbf {\bibinfo
  {volume} {997}} (\bibinfo {year} {2023})}\BibitemShut {NoStop}%
\bibitem [{\citenamefont {Karthein}\ \emph {et~al.}(2023)\citenamefont
  {Karthein}, \citenamefont {Ricketts}, \citenamefont {Ruiz}, \citenamefont
  {Billowes}, \citenamefont {Binnersley}, \citenamefont {Cocolios},
  \citenamefont {Dobaczewski}, \citenamefont {Farooq-Smith}, \citenamefont
  {Flanagan}, \citenamefont {Georgiev} \emph
  {et~al.}}]{karthein2023electromagnetic}%
  \BibitemOpen
  \bibfield  {author} {\bibinfo {author} {\bibfnamefont {J.}~\bibnamefont
  {Karthein}}, \bibinfo {author} {\bibfnamefont {C.}~\bibnamefont {Ricketts}},
  \bibinfo {author} {\bibfnamefont {R.}~\bibnamefont {Ruiz}}, \bibinfo {author}
  {\bibfnamefont {J.}~\bibnamefont {Billowes}}, \bibinfo {author}
  {\bibfnamefont {C.}~\bibnamefont {Binnersley}}, \bibinfo {author}
  {\bibfnamefont {T.}~\bibnamefont {Cocolios}}, \bibinfo {author}
  {\bibfnamefont {J.}~\bibnamefont {Dobaczewski}}, \bibinfo {author}
  {\bibfnamefont {G.}~\bibnamefont {Farooq-Smith}}, \bibinfo {author}
  {\bibfnamefont {K.}~\bibnamefont {Flanagan}}, \bibinfo {author}
  {\bibfnamefont {G.}~\bibnamefont {Georgiev}},  \emph {et~al.},\ }\href@noop
  {} {\bibfield  {journal} {\bibinfo  {journal} {arXiv preprint
  arXiv:2310.15093}\ } (\bibinfo {year} {2023})}\BibitemShut {NoStop}%
\bibitem [{\citenamefont {Bohr}(1976)}]{1976BohrRMP}%
  \BibitemOpen
  \bibfield  {author} {\bibinfo {author} {\bibfnamefont {A.}~\bibnamefont
  {Bohr}},\ }\href {\doibase 10.1103/RevModPhys.48.365} {\bibfield  {journal}
  {\bibinfo  {journal} {Rev. Mod. Phys.}\ }\textbf {\bibinfo {volume} {48}},\
  \bibinfo {pages} {365} (\bibinfo {year} {1976})}\BibitemShut {NoStop}%
\bibitem [{\citenamefont {Mottelson}(1976)}]{1976MottelsonRMP}%
  \BibitemOpen
  \bibfield  {author} {\bibinfo {author} {\bibfnamefont {B.}~\bibnamefont
  {Mottelson}},\ }\href {\doibase 10.1103/RevModPhys.48.375} {\bibfield
  {journal} {\bibinfo  {journal} {Rev. Mod. Phys.}\ }\textbf {\bibinfo {volume}
  {48}},\ \bibinfo {pages} {375} (\bibinfo {year} {1976})}\BibitemShut
  {NoStop}%
\bibitem [{\citenamefont {Naqvi}\ \emph {et~al.}(1995)\citenamefont {Naqvi},
  \citenamefont {Bahri}, \citenamefont {Troltenier}, \citenamefont {Draayer},\
  and\ \citenamefont {Faessler}}]{1995NaqviZPhysA}%
  \BibitemOpen
  \bibfield  {author} {\bibinfo {author} {\bibfnamefont {H.~A.}\ \bibnamefont
  {Naqvi}}, \bibinfo {author} {\bibfnamefont {C.}~\bibnamefont {Bahri}},
  \bibinfo {author} {\bibfnamefont {D.}~\bibnamefont {Troltenier}}, \bibinfo
  {author} {\bibfnamefont {J.~P.}\ \bibnamefont {Draayer}}, \ and\ \bibinfo
  {author} {\bibfnamefont {A.}~\bibnamefont {Faessler}},\ }\href {\doibase
  10.1007/BF01290907} {\bibfield  {journal} {\bibinfo  {journal} {Zeitschrift
  f{\"u}r Physik A Hadrons and Nuclei}\ }\textbf {\bibinfo {volume} {351}},\
  \bibinfo {pages} {259} (\bibinfo {year} {1995})}\BibitemShut {NoStop}%
\bibitem [{\citenamefont {Rowe}\ \emph {et~al.}(2016)\citenamefont {Rowe},
  \citenamefont {McCoy},\ and\ \citenamefont {Caprio}}]{2016RowePhysScripta}%
  \BibitemOpen
  \bibfield  {author} {\bibinfo {author} {\bibfnamefont {D.~J.}\ \bibnamefont
  {Rowe}}, \bibinfo {author} {\bibfnamefont {A.~E.}\ \bibnamefont {McCoy}}, \
  and\ \bibinfo {author} {\bibfnamefont {M.~A.}\ \bibnamefont {Caprio}},\
  }\href {\doibase 10.1088/0031-8949/91/3/033003} {\bibfield  {journal}
  {\bibinfo  {journal} {Physica Scripta}\ }\textbf {\bibinfo {volume} {91}},\
  \bibinfo {pages} {033003} (\bibinfo {year} {2016})}\BibitemShut {NoStop}%
\bibitem [{\citenamefont {Raghavan}(1989)}]{1989RaghavanTables}%
  \BibitemOpen
  \bibfield  {author} {\bibinfo {author} {\bibfnamefont {P.}~\bibnamefont
  {Raghavan}},\ }\href {\doibase https://doi.org/10.1016/0092-640X(89)90008-9}
  {\bibfield  {journal} {\bibinfo  {journal} {Atomic Data and Nuclear Data
  Tables}\ }\textbf {\bibinfo {volume} {42}},\ \bibinfo {pages} {189} (\bibinfo
  {year} {1989})}\BibitemShut {NoStop}%
\bibitem [{\citenamefont {Hergert}(2020)}]{hergert2020guided}%
  \BibitemOpen
  \bibfield  {author} {\bibinfo {author} {\bibfnamefont {H.}~\bibnamefont
  {Hergert}},\ }\href@noop {} {\bibfield  {journal} {\bibinfo  {journal}
  {Frontiers in Physics}\ }\textbf {\bibinfo {volume} {8}},\ \bibinfo {pages}
  {379} (\bibinfo {year} {2020})}\BibitemShut {NoStop}%
\bibitem [{\citenamefont {Zare}(1988)}]{1988ZareBook}%
  \BibitemOpen
  \bibfield  {author} {\bibinfo {author} {\bibfnamefont {R.}~\bibnamefont
  {Zare}},\ }\href@noop {} {\emph {\bibinfo {title} {Angular Momentum:
  Understanding Spatial Aspects in Chemistry and Physics}}},\ 99-0102851-5\
  (\bibinfo  {publisher} {Wiley},\ \bibinfo {year} {1988})\BibitemShut
  {NoStop}%
\bibitem [{\citenamefont {Micu}(1968)}]{1968MicuNucPhysA}%
  \BibitemOpen
  \bibfield  {author} {\bibinfo {author} {\bibfnamefont {M.}~\bibnamefont
  {Micu}},\ }\href {\doibase https://doi.org/10.1016/0375-9474(68)90895-6}
  {\bibfield  {journal} {\bibinfo  {journal} {Nuclear Physics A}\ }\textbf
  {\bibinfo {volume} {113}},\ \bibinfo {pages} {215} (\bibinfo {year}
  {1968})}\BibitemShut {NoStop}%
\bibitem [{\citenamefont {Raynal}(1979)}]{1979RaynalJMatPhys}%
  \BibitemOpen
  \bibfield  {author} {\bibinfo {author} {\bibfnamefont {J.}~\bibnamefont
  {Raynal}},\ }\href {\doibase 10.1063/1.524047} {\bibfield  {journal}
  {\bibinfo  {journal} {Journal of Mathematical Physics}\ }\textbf {\bibinfo
  {volume} {20}},\ \bibinfo {pages} {2398} (\bibinfo {year}
  {1979})}\BibitemShut {NoStop}%
\bibitem [{\citenamefont {Pain}(2020)}]{2020PainEPJA}%
  \BibitemOpen
  \bibfield  {author} {\bibinfo {author} {\bibfnamefont {J.-C.}\ \bibnamefont
  {Pain}},\ }\href {\doibase 10.1140/epja/s10050-020-00303-9} {\bibfield
  {journal} {\bibinfo  {journal} {The European Physical Journal A}\ }\textbf
  {\bibinfo {volume} {56}},\ \bibinfo {pages} {296} (\bibinfo {year}
  {2020})}\BibitemShut {NoStop}%
\bibitem [{\citenamefont {{International Atomic Energy
  Agency}}(2023)}]{IAEAGammaRayData}%
  \BibitemOpen
  \bibfield  {author} {\bibinfo {author} {\bibnamefont {{International Atomic
  Energy Agency}}},\ }\href@noop {} {\enquote {\bibinfo {title} {Gamma-ray
  energies database},}\ }\bibinfo {howpublished}
  {\url{https://www-nds.iaea.org/}} (\bibinfo {year} {2023}),\ \bibinfo {note}
  {accessed: 20-12-2023}\BibitemShut {NoStop}%
\bibitem [{\citenamefont {Doser}(2022)}]{doser2022antiprotonic}%
  \BibitemOpen
  \bibfield  {author} {\bibinfo {author} {\bibfnamefont {M.}~\bibnamefont
  {Doser}},\ }\href@noop {} {\bibfield  {journal} {\bibinfo  {journal}
  {Progress in Particle and Nuclear Physics}\ }\textbf {\bibinfo {volume}
  {125}},\ \bibinfo {pages} {103964} (\bibinfo {year} {2022})}\BibitemShut
  {NoStop}%
\bibitem [{\citenamefont {Rodin}\ \emph {et~al.}(2022)\citenamefont {Rodin},
  \citenamefont {Doser}, \citenamefont {Khatri}, \citenamefont {Cerchiari},
  \citenamefont {Farricker}, \citenamefont {Welsch},\ and\ \citenamefont
  {Kornakov}}]{rodin2022low}%
  \BibitemOpen
  \bibfield  {author} {\bibinfo {author} {\bibfnamefont {V.}~\bibnamefont
  {Rodin}}, \bibinfo {author} {\bibfnamefont {M.}~\bibnamefont {Doser}},
  \bibinfo {author} {\bibfnamefont {G.}~\bibnamefont {Khatri}}, \bibinfo
  {author} {\bibfnamefont {G.}~\bibnamefont {Cerchiari}}, \bibinfo {author}
  {\bibfnamefont {A.}~\bibnamefont {Farricker}}, \bibinfo {author}
  {\bibfnamefont {C.}~\bibnamefont {Welsch}}, \ and\ \bibinfo {author}
  {\bibfnamefont {G.}~\bibnamefont {Kornakov}},\ }\href@noop {} {\bibfield
  {journal} {\bibinfo  {journal} {JACoW IPAC}\ }\textbf {\bibinfo {volume}
  {2022}},\ \bibinfo {pages} {2126} (\bibinfo {year} {2022})}\BibitemShut
  {NoStop}%
\bibitem [{\citenamefont {Kornakov}\ \emph {et~al.}(2023)\citenamefont
  {Kornakov}, \citenamefont {Cerchiari}, \citenamefont {Zieli{\'n}ski},
  \citenamefont {Lappo}, \citenamefont {Sadowski},\ and\ \citenamefont
  {Doser}}]{kornakov2023synthesis}%
  \BibitemOpen
  \bibfield  {author} {\bibinfo {author} {\bibfnamefont {G.}~\bibnamefont
  {Kornakov}}, \bibinfo {author} {\bibfnamefont {G.}~\bibnamefont {Cerchiari}},
  \bibinfo {author} {\bibfnamefont {J.}~\bibnamefont {Zieli{\'n}ski}}, \bibinfo
  {author} {\bibfnamefont {L.}~\bibnamefont {Lappo}}, \bibinfo {author}
  {\bibfnamefont {G.}~\bibnamefont {Sadowski}}, \ and\ \bibinfo {author}
  {\bibfnamefont {M.}~\bibnamefont {Doser}},\ }\href@noop {} {\bibfield
  {journal} {\bibinfo  {journal} {Physical Review C}\ }\textbf {\bibinfo
  {volume} {107}},\ \bibinfo {pages} {034314} (\bibinfo {year}
  {2023})}\BibitemShut {NoStop}%
\bibitem [{\citenamefont {Caravita}(2023)}]{AEgISReport2022}%
  \BibitemOpen
  \bibfield  {author} {\bibinfo {author} {\bibfnamefont {R.}~\bibnamefont
  {Caravita}} (\bibinfo {collaboration} {AEgIS}),\ }\href
  {https://cds.cern.ch/record/2846698} {\emph {\bibinfo {title} {{AEgIS/AD-6
  annual report 2022}}}},\ \bibinfo {type} {Tech. Rep.}\ (\bibinfo
  {institution} {CERN},\ \bibinfo {address} {Geneva},\ \bibinfo {year}
  {2023})\BibitemShut {NoStop}%
\bibitem [{\citenamefont {Caravita}(2024)}]{AEgISReport2023}%
  \BibitemOpen
  \bibfield  {author} {\bibinfo {author} {\bibfnamefont {R.}~\bibnamefont
  {Caravita}} (\bibinfo {collaboration} {AEgIS}),\ }\href
  {https://cds.cern.ch/record/2887577} {\emph {\bibinfo {title} {{AEgIS/AD-6
  annual report 2023}}}},\ \bibinfo {type} {Tech. Rep.}\ (\bibinfo
  {institution} {CERN},\ \bibinfo {address} {Geneva},\ \bibinfo {year}
  {2024})\BibitemShut {NoStop}%
\bibitem [{\citenamefont {Aumann}\ \emph {et~al.}(2022)\citenamefont {Aumann},
  \citenamefont {Bartmann}, \citenamefont {Boine-Frankenheim}, \citenamefont
  {Bouvard}, \citenamefont {Broche}, \citenamefont {Butin}, \citenamefont
  {Calvet}, \citenamefont {Carbonell}, \citenamefont {Chiggiato}, \citenamefont
  {De~Gersem} \emph {et~al.}}]{aumann2022puma}%
  \BibitemOpen
  \bibfield  {author} {\bibinfo {author} {\bibfnamefont {T.}~\bibnamefont
  {Aumann}}, \bibinfo {author} {\bibfnamefont {W.}~\bibnamefont {Bartmann}},
  \bibinfo {author} {\bibfnamefont {O.}~\bibnamefont {Boine-Frankenheim}},
  \bibinfo {author} {\bibfnamefont {A.}~\bibnamefont {Bouvard}}, \bibinfo
  {author} {\bibfnamefont {A.}~\bibnamefont {Broche}}, \bibinfo {author}
  {\bibfnamefont {F.}~\bibnamefont {Butin}}, \bibinfo {author} {\bibfnamefont
  {D.}~\bibnamefont {Calvet}}, \bibinfo {author} {\bibfnamefont
  {J.}~\bibnamefont {Carbonell}}, \bibinfo {author} {\bibfnamefont
  {P.}~\bibnamefont {Chiggiato}}, \bibinfo {author} {\bibfnamefont
  {H.}~\bibnamefont {De~Gersem}},  \emph {et~al.},\ }\href@noop {} {\bibfield
  {journal} {\bibinfo  {journal} {The European Physical Journal A}\ }\textbf
  {\bibinfo {volume} {58}},\ \bibinfo {pages} {88} (\bibinfo {year}
  {2022})}\BibitemShut {NoStop}%
\end{thebibliography}%

\end{document}